\documentclass[showpacs,aps,prl,twocolumn,superscriptaddress]{revtex4}

\usepackage{graphicx}% Include figure files

\begin{document}

%\preprint{}

\title{Electron localization near Mott transition in organic superconductor $\kappa$-(BEDT-TTF)$_{2}$Cu[N(CN)$_{2}]$Br}



\author{K. Sano}
\affiliation{Institute for Materials Research, Tohoku University, Sendai 980-8577, Japan}

\author{T. Sasaki}
\email{takahiko@imr.tohoku.ac.jp}
\affiliation{Institute for Materials Research, Tohoku University, Sendai 980-8577, Japan}
\affiliation{Japan Science and Technology Agency, CREST, Tokyo 102-0075, Japan}

\author{N. Yoneyama}
\thanks{Present address: University of Yamanashi, Japan}
\affiliation{Institute for Materials Research, Tohoku University, Sendai 980-8577, Japan}
\affiliation{Japan Science and Technology Agency, CREST, Tokyo 102-0075, Japan}

\author{N. Kobayashi}
\affiliation{Institute for Materials Research, Tohoku University, Sendai 980-8577, Japan}


\date{\today}

\begin{abstract}
The effect of disorder on the electronic properties near the Mott transition is studied in an organic superconductor $\kappa$-(BEDT-TTF)$_{2}$Cu[N(CN)$_{2}$]Br, which is systematically irradiated by X-ray.
We observe that X-ray irradiation causes Anderson-type electron localization due to molecular disorder.  
The resistivity at low temperatures demonstrates variable range hopping conduction with Coulomb interaction.  
The experimental results show clearly that the electron localization by disorder is enhanced by the Coulomb interaction near the Mott transition.

\end{abstract}

%\keywords{}

\maketitle


%Introduction
Metal-insulator (MI) transitions are of considerable importance for strongly correlated electron systems.
Among the various types of the MI transitions, the Mott transition due to electron-electron (e-e) interactions is one of the most attractive phenomena \cite{Imada}.
A Mott insulator derives from the large on-site Coulomb energy with respect to the bandwidth. 
The electrons in the Mott insulator are localized on the individual sites to minimize the mutual Coulomb repulsion, which results in the opening of a charge gap at the Fermi level. 
Another way of the electron localization originates from the interference of the electron wave functions due to randomness.  
This is the Anderson insulator derived by introducing disorder into the material \cite{Kramer}.
In contrast to Mott insulators, in Anderson insulators, there is no opening of a gap in the density of states in principle.  
Since the randomness in the correlated electron system is essentially important in real materials, systematic studies of disorder effects are desired in systems nearby a Mott transition for understanding their physical properties.
The e-e interaction effects in disordered systems have been considered in the weak localization (WL) effect of electrons \cite{Altshuler,Fukuyama}.
Recently, several theoretical studies have been performed in consideration of the Mott transition \cite{Shinaoka1,Shinaoka2,Byczuk}, but there has been few experimental approaches so far \cite{Kim} because of limited suitable materials and ways of introducing disorder.  

Organic charge-transfer salts based on a donor molecule bis(ethylenedithio)-tetrathiafulvalene (abbreviated as BEDT-TTF) have been recognized as highly correlated electron systems \cite{Powell}.  
Among them, $\kappa$-(BEDT-TTF)$_{2}$$X$, where $X$ is anion molecule, has attracted considerable attention as a bandwidth-controlled Mott transition system \cite{Kanoda}. 
One can control the strength of electron correlation relative to the bandwidth by small pressure \cite{Kagawa} or partial molecule substitution \cite{Yoneyama}, which leads to a Mott insulator - metal transition. 
Recently, X-ray irradiation effects in $\kappa$-(BEDT-TTF)$_{2}$$X$ have been examined \cite{Analytis,Sasaki1,Sasaki2}.  
It has been known that X-ray irradiation causes molecular disorder and this remains permanently in organic materials \cite{Zuppirol}.
In case of the superconductor $\kappa$-(BEDT-TTF)$_{2}$Cu(NCS)$_{2}$ (hereafter $\kappa$-NCS), the molecular disorder causes an increase of the residual resistivity and then suppresses the superconductivity \cite{Analytis}.
On the other hand in case of the Mott insulator $\kappa$-(BEDT-TTF)$_{2}$Cu[N(CN)$_{2}$]Cl (hereafter $\kappa$-Cl), the small potential modulation by disorder is accompanied by a small shift of the band-filling in addition \cite{Sasaki2}.  

In this Letter, we report the disorder effect on the electronic properties in the organic superconductor $\kappa$-(BEDT-TTF)$_{2}$Cu[N(CN)$_{2}$]Br (hereafter $\kappa$-Br) which is located closer to the Mott transition on the metallic side compared to $\kappa$-NCS.  
We demonstrate that the stronger electron correlation upon approaching to the Mott transition enhances Anderson-type electron localization due to disorder introduced by X-ray irradiation.

%Experiments

Single crystals of $\kappa$-Br were grown by a standard electrochemical oxidation method.  
The in-plane resistivity was measured by the dc four-terminal method with the current parallel to the $a$-$c$ plane. 
The measurements under pressure were performed by using a standard piston cylinder clamp cell.
The samples were irradiated at 300 K using a nonfiltered tungsten target at 40 kV and 20 mA.  
The dose rate in the present experimental conditions was approximately 0.5 MGy/hour \cite{Sasaki1}. 
After each step of the irradiation, the temperature dependence of the resistivity was measured at the same cooling rate of $- 0.4$ K/min.
The irradiation time $t^{\rm irr}$ was the sum of multiple exposures at 300 K.
We examined eight samples.
The resistivity values before X-ray irradiation vary within approximately an order of magnitude. 
It may originate from the inaccuracy of the sample and contact dimensions and unexpected inhomogeneous current distribution in the layered material. 
However the qualitative properties and even quantitative parameters obtained in response to X-ray irradiation have no significant sample-to-sample variation. 

%Results and Discussion

\begin{figure}
\includegraphics[viewport=2cm 9cm 17cm 22.5cm,clip,width=1.0\linewidth]{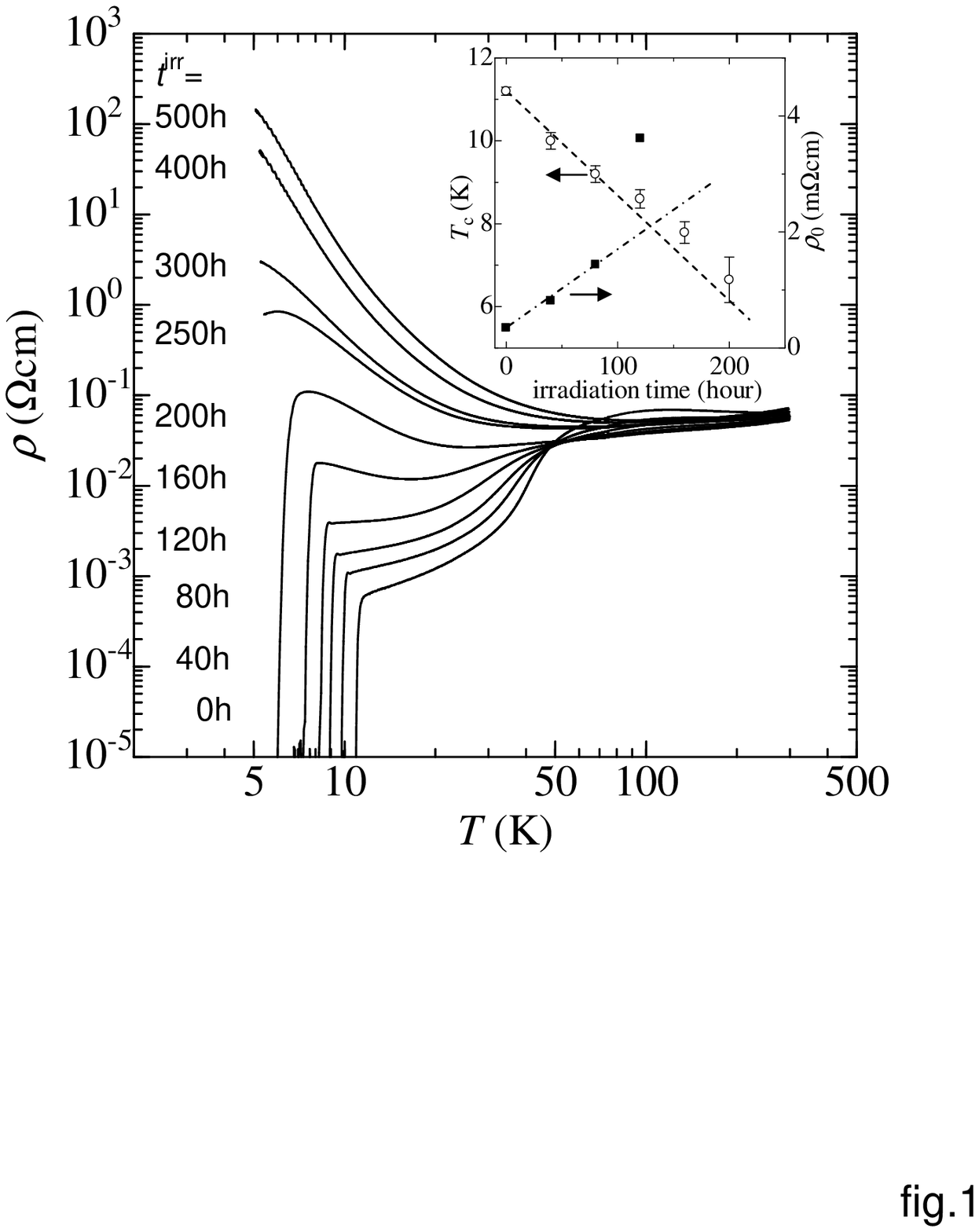}
\caption{Temperature dependence of the in-plane resistivity of $\kappa$-(BEDT-TTF)$_{2}$Cu[N(CN)$_{2}$]Br irradiated by X-ray. The inset shows the changes of $T_{\rm c}$ and $\rho_{0}$ with X-ray irradiation time $t^{\rm irr}$. The mid point of the resistivity transition is defined as $T_{\rm c}$ and the bars indicate onset and offset of the transition. The lines are guide for eyes.}
\end{figure}

Figure 1 shows the temperature dependence of the resistivity of $\kappa$-Br irradiated by X-ray.  
Before irradiation, the resistivity shows the characteristic behavior reported previously \cite{Lang}: a broad resistivity hump at 100 K, a crossover around $T^{*}$ from a bad to good metallic state at low temperature where the resistivity follows $\rho(T) = \rho_{0} + $A$T^{2}$ and the superconducting transition at $T_{\rm c} \simeq$ 11 K.  

X-ray irradiation to the sample changes the behavior of the resistivity drastically.  
In the low irradiation dose for $t^{\rm irr} <$ 100 h, the residual resistivity $\rho_{0}$ increases and $T_{\rm c}$ decreases almost linearly with $t^{\rm irr}$ as shown in the inset.  
Concurrently, the characteristic hump structure around 100 K is suppressed and the $\rho(T)$ curves intersect a single point at $T =$ 50 K and $\rho \simeq$ 30 m$\Omega$cm.  
These changes have been similarly observed in $\kappa$-NCS \cite{Analytis}.  
In contrast to the previous results of $\kappa$-NCS, however, further irradiation causes not only the increase of disorder scattering but also electron localization.  
The values of $T_{\rm c}$ and $\rho_{0}$ start to deviate from the linear dependence at $t^{\rm irr} >$ 100 h. 
It is noted that $\rho$($T$) at $t^{\rm irr} =$ 200 h shows almost temperature independent behavior at the same resistivity range of the intersection point.
At $t^{\rm irr} >$ 250 h, $\rho$($T$) curves show insulating behavior at low temperature and the curves do not cross the intersection point.
At $t^{\rm irr} =$ 500 h at last, the resistivity at 4 K becomes more than five orders of magnitude larger than $\rho_{0}$ before irradiation.
The change of the resistivity clearly indicates that the observed MI transition is induced by disorder introduced by X-ray irradiation.  
Although the suppression of superconductivity \cite{Analytis} and its relation to the insulating behavior are important issues, we focus on the insulating behavior which results from an enhancement of Anderson-type localization with Coulomb interaction nearby the Mott transition.  

\begin{figure}
\includegraphics[viewport=1cm 10cm 20cm 24cm,clip,width=1.0\linewidth]{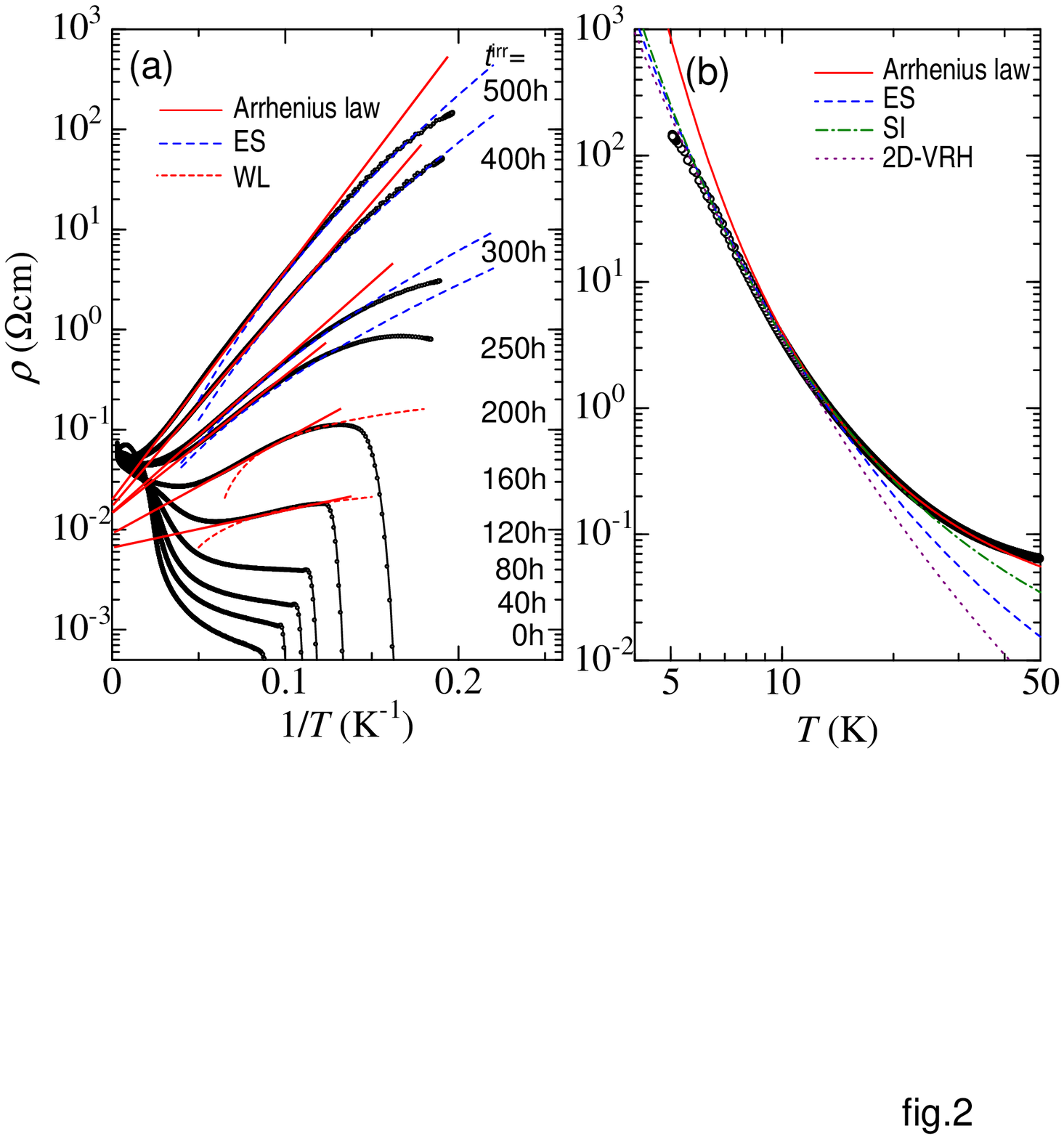}
\caption{(color online) (a) Arrhenius plot of the resistivity of $\kappa$-(BEDT-TTF)$_{2}$Cu[N(CN)$_{2}$]Br irradiated by X-ray. Red solid lines indicate an Arrhenius-law type of $\rho(T)$. Blue dashed curves are the fitting results for the model by Efros and Shklovskii (ES). Red dotted curves represent the $\log(T)$ dependence for weak localization (WL). (b) Fitting curves for $\rho(T)$ at $t^{\rm irr} =$ 500 h for the model of the Arrhenius law, by Efros and Shklovskii (ES), by Shinaoka and Imada (SI) and of the variable range hopping in two dimensions (2D-VRH). }
\end{figure}

Figure 2(a) shows an Arrhenius plot of the same data as in Fig. 1. 
In the temperature range 10 K $< T <$ 40 K, the resistivity follows well the Arrhenius law described as $\rho(T) = \rho_{a}\exp\left(E_{a}/k_{\rm B}T\right)$, where $\rho_{a}$ is the resistivity extrapolated to the high-temperature limit and $E_{a}$ is the characteristic energy.  
The Arrhenius-type temperature dependence usually appears in band insulators (semiconductors) and $E_{a}$ corresponds to the band gap energy.  
The present results, however, do not originate from a thermally activation across a band-gap, but electron hopping between nearest-neighbor localized sites in the disordered system at high temperature \cite{Mott}.  
This is demonstrated by the downward deviation of the resistivity from the Arrhenius law at lower temperature, which is not expected in a band-gap insulator.  
The energy $E_{a}$ in the hopping conduction corresponds to the degree of the randomness.  
As shown in the lower part of Fig. 3, $E_{a}$ at $t^{\rm irr} >$ 200 h increases linearly with $t^{\rm irr}$ and the extrapolation to lower irradiation times suggests an intersect with the origin.  
At $t^{\rm irr} <$ 200 h (open triangles), there are some deviations from the linear dependence.  
This may be due to the intermediate behavior between insulator and metal.
Actually, $\rho(T)$ shows the $\log(T)$ dependence in the intermediate irradiation at $t^{\rm irr} =$ 160 -- 200 h as shown in Fig. 2(a), which suggests the WL behavior in the metallic side for the MI transition.  
The extrapolated value of $\rho_{a} \simeq$ 20 m$\Omega$cm in the high-temperature limit is reasonably close to the nearly temperature-independent resistivity at $t^{\rm irr} =$ 200 h.  
Here we estimate the corresponding sheet resistance of 130 k$\Omega$ from $\rho_{a}$ assuming the layer distance of 1.5 nm.  
This value is comparable to the inverse of the minimum conductivity in two dimensions \cite{Ando}, $\sigma_{\rm min} \simeq 0.1e^{2}/\hbar \simeq 2.4 \times 10^{-5}$ S, and then 1/$\sigma_{\rm min} \simeq$ 40 k$\Omega$.  
The fairly good correspondence even though rough estimation could suggest that the observations of $\rho_{a}$ and temperature independent resistivity showed the features of the critical resistivity for the MI transition induced by disorder. 
In the meantime, the resistivity ($\sim$ 30m${\Omega}$cm) at the intersection point of $\rho(T)$ curves also closes to $\rho_{\rm a}$.  
At present, however, we have no definite explanation for this relation although the origin of the intersection point may be concerned with the MI transition. 

The resistivity below about 10 K which deviates from the Arrhenius law can be fitted to the following function, $\rho(T) = \rho_{\rm h}\exp\left[\left(T_{0}/T\right)^{n}\right]$, where $n =$ 1/3 is expected for variable range hopping (VRH) in two dimensions, and 1/4 in three dimensions \cite{Mott}.
The case where $n = 1/2$ has been known to appear in localized states in the presence of long-range Coulomb repulsive interactions, which has been proposed by Efros and Shklovskii (ES) \cite{Efros}.  
This being the case leads to the formation of a Coulomb gap.  
The fitting results are fairly well with all of $n =$ 1/2, 1/3 and 1/4 on an equality, for example, as shown by dashed curves for the ES model with $n =$ 1/2 in Fig. 2(a).  
Therefore we could not judge the best suitable fractional number $n$ from the curve fit procedure.  
Considering the strong two dimensionality and the strong electron correlation in this material, however, VRH in two dimensions (2D-VRH) or the ES model should be valid.  

We examined the insulating behavior in $\rho(T)$ with considering WL including e-e interactions \cite{Altshuler,Fukuyama}, in addition to a model including short-range Coulomb interactions with disorder by Shinaoka and Imada (SI) \cite{Shinaoka1,Shinaoka2}.  
The former, however, does not fit to the observed exponential increase of $\rho(T)$ since it modifies only the prefactor of the $\log(T)$ term in $\rho(T)$.  
Figure 2(b) shows the fitting results for the resistivity at $t^{\rm irr} =$ 500 h.  
The curve of SI is calculated by the following equation \cite{Shinaoka2},
\begin{equation}
\rho(T) = \rho_{\rm SI}\exp\left\{c_{0}\frac{\exp\left[-c_{1}\left|\log(k_{\rm B}T)^{1/d}\right|\right]}{k_{\rm B}T}\right\},
\end{equation}
where $\rho_{\rm SI}$, $c_{0}$ and $c_{1}$ are fitting parameters and $d$ is the spacial dimension and here $d =$ 2.  
At low temperature, all of ES, 2D-VRH and SI coincide with each other.  
On the other hand, the Arrhenius law is applicable at high temperatures. 
In the intermediate region, SI may be a better description than ES and 2D-VRH.
The crossover from SI at higher temperatures to ES or 2D-VRH at lower temperatures is simply understood by changing the range of Coulomb interactions from short-range to long-range due to weak electron screening capability with decreasing temperature.  

\begin{figure}
\includegraphics[viewport=2.5cm 8cm 18cm 23.5cm,clip,width=1.0\linewidth]{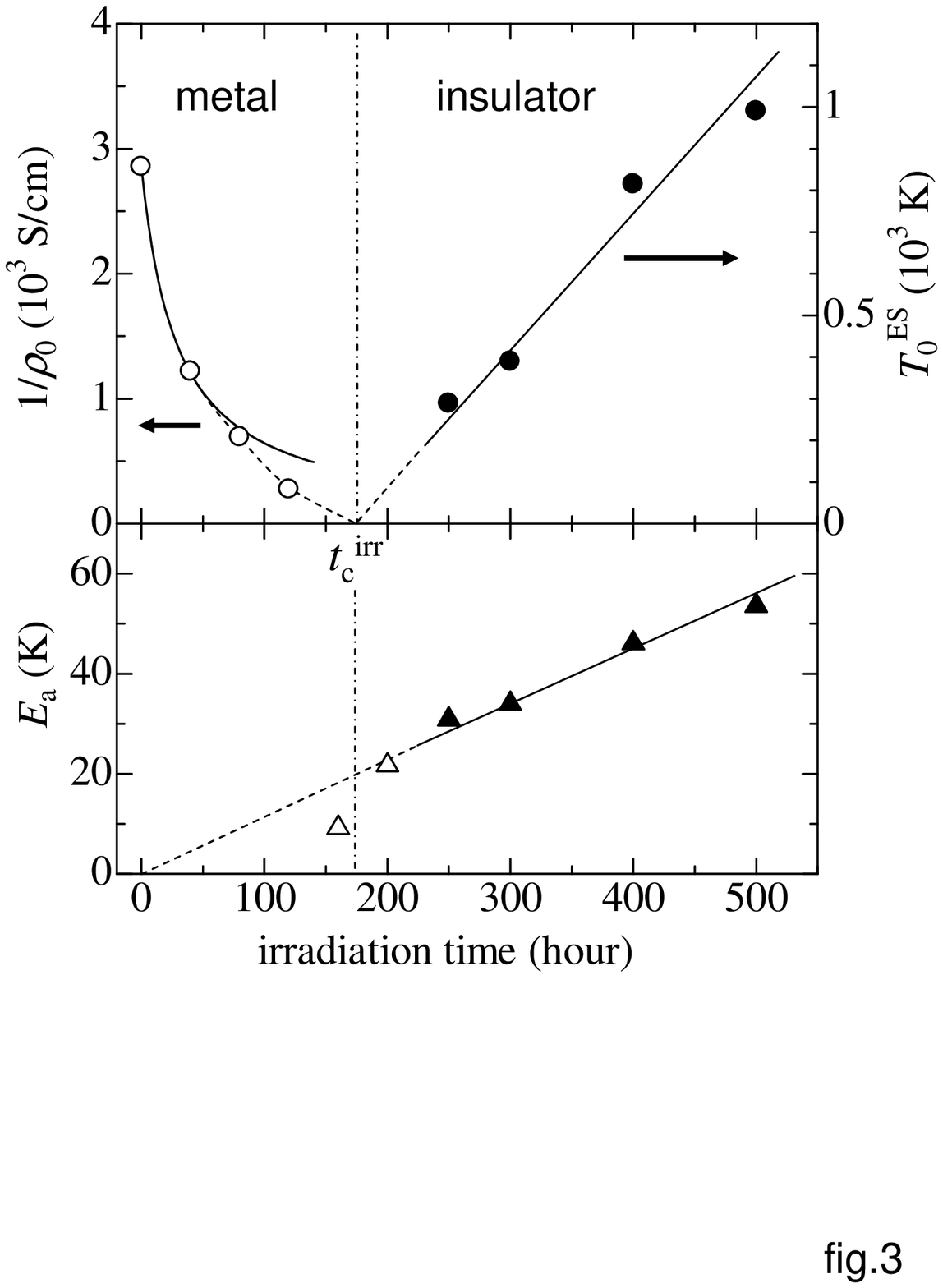}
\caption{Irradiation time dependence of the residual resistivity $\rho_{0}$, $T_{0}^{\rm ES}$ in the ES model and $E_{\rm a}$ in the Arrhenius law. Solid curve for $1/\rho_{0}$ represents the the linear dependence of $\rho_{0}(t^{\rm irr})$.}
\end{figure}

Next, we discuss the critical behavior of the MI transition induced by disorder.  
The critical property appears as a divergence of the localization length $\xi$ for variable parameters.
The irradiation time $t^{\rm irr}$ is the parameter in the present case.
The temperature coefficient $T_{0}$ is related to $\xi$ as $k_{\rm B}T_{0}^{\rm ES} \simeq e^{2}/\kappa\xi$ in the case of ES \cite{Efros}, where $\kappa$ is the dielectric constant. 
Figure 3 shows the critical behavior of some parameters for the MI transition as a function of $t^{\rm irr}$. 
In the insulator region, $T_{0}^{\rm ES}$ increases almost linearly with $t^{\rm irr}$. 
One could obtain a critical irradiation time $t_{\rm c}^{\rm irr}$ of 150 -- 200 h from the extrapolation to $T_{0}^{\rm ES}$($t^{\rm irr}$) $=$ 0, where the localization length diverges as $\xi \propto (t^{\rm irr} - t^{\rm irr}_{\rm c})^{-1}$.  
The value of $\xi$ could be estimated to be in the range of 0.2 - 2 nm at $t^{\rm irr} =$ 500 h if we assumed $\kappa = 10 - 100$ though the dielectric properties of the present irradiated sample have not been measured so far.
The estimated value is reasonably consistent with the site hopping scenario of ES and VRH in comparison to the mean distance ($\sim$ 0.3 nm) of BEDT-TTF molecules. 
On the other hand, the inverse of $\rho_{0}$ is plotted in the metal region and the solid curve represents the linear dependence for $t^{\rm irr}$ as shown in the inset of Fig. 1.  
The extrapolation of 1/$\rho_{0}$ goes to zero at $t^{\rm irr}_{\rm c}$. 
The critical behavior for $\xi$ and $\rho_{0}$ with a convergence at a single $t_{\rm c}^{\rm irr}$ from both insulator and metal sides strongly supports that there is a critical disorder for the MI transition.  
The obtained $t_{\rm c}^{\rm irr} \simeq$ 150 -- 200 h ensures also the critical resistivity for the MI transition as mentioned above.  
It is interestingly noted that the value of $E_{a}$ does not show critical behavior at $t^{\rm irr}_{\rm c}$ because $E_{a}$ may simply represent the randomness.  

\begin{figure}
\includegraphics[viewport=1cm 8cm 19cm 21.5cm,clip,width=0.9\linewidth]{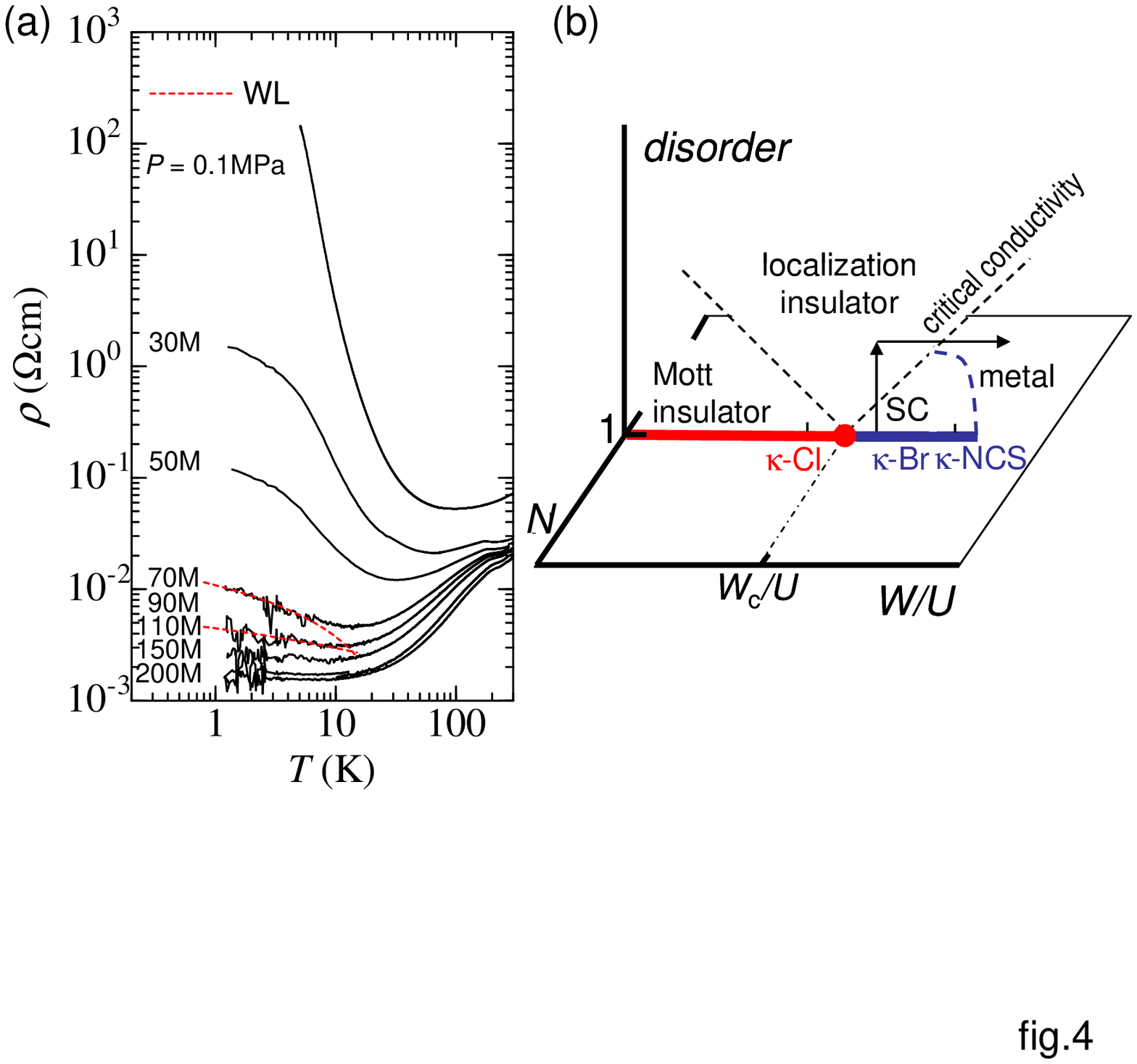}
\caption{(color online) (a) Temperature dependence of the resistivity in the 500 h irradiated sample under pressures. The pressure values are corrected by the effect of the solidification of the pressure medium. The dotted curves represent the WL behavior. (b) Schematic phase diagram for band width $W$, carrier number $N$ and disorder. The vertical and horizontal arrows represent the experimental processes of the X-ray irradiation for introducing disorder and then the application of pressure to increase $W$, respectively. }
\end{figure}

It is important to examine the pressure effect in order to consider the relation between electron correlations and disorder.  
Figure 4(a) shows the pressure effect for the resistivity after 500 h X-ray irradiation. 
With increasing pressure, the resistivity decreases and the insulating behavior becomes weaker.  
After going through the resistivity level of the same order as the critical resistivity for the MI transition, the WL behavior, $\rho(T) \propto -\log(T)$, is found below approximately 10 K at 70 -- 90 MPa and then a metallic behavior is restored at 150 -- 200 MPa.  
Superconductivity, however, is not observed down to 1.5 K. 
In this process, pressure broadens the band width $W$ in comparison to the on-site Coulomb energy $U$ resulting in weaker electron correlations. 
At the same time, the amount of disorder is not influenced essentially by pressure.
A schematic electronic phase diagram with the axes of the carrier number per site $N$, the band width $W$ and disorder is depicted in Fig. 4(b).  
On the line of $N =$ 1, corresponding to the half-filling of the conduction band, the Mott insulator - metal transition takes place at the critical bandwidth $W_{\rm c}$.  
The present $\kappa$-Br is located nearby the Mott transition on the metallic side.  
The introduction of disorder changes the metal to an Anderson-type localization insulator after crossing a critical disorder value determined by the minimum conductivity.  
Then the broader $W$ induced by pressure brings the insulator back to the metallic state at the same minimum conductivity. 
In the metallic side for the MI transition, the WL behavior appears in $\rho(T)$.  
This schematic phase diagram for disorder represents the present experimental observation well \cite{Sasaki3}.  
In addition, this phase diagram may respond to the reason why the other organic superconductor $\kappa$-NCS does not show insulating behavior but only show increasing the residual resistivity although almost the same level of disorder is introduced by the same way of X-ray irradiation. 
This may be due to the wider bandwidth of $\kappa$-NCS than that of $\kappa$-Br.  
In that case, the electron correlation is not strong enough to cause localization of carriers by the same amount of disorder in comparison with $\kappa$-Br. 
In $\kappa$-Cl, the Mott gap is suppressed by the local disorder effect which induces small shift of band-filling \cite{Sasaki2,Yoneyama2}, and then the Mott insulator changes to the localization one due to disorder.

In summary, we investigated the disorder effect on the electron localization in the X-ray irradiated organic superconductor $\kappa$-Br which is located nearby the Mott transition.  
We demonstrated that the disorder-induced localization of the electrons is enhanced by electron correlations.  

%acknowledgments
The authors thank H. Shinaoka, M. Imada, M. Lang, J. M\"uller, and T. Nakamura for valuable discussions.
This work was partly supported by a Grant-in-Aid for Scientific Research (No. 20340085 and 21110504) from MEXT and JSPS, Japan.


\end{document}